# Production of Indium Doped Zinc Oxide Thin Films by Pulsed Laser Ablation

V. Savchuk[1*], B. Kotlyarchuk[1], M. Oszwaldowski[2]

[1] *Pidstryhach Institute for Applied Problems of Mechanics and Mathematics, National Academy of Sciences of Ukraine, 3B Naukova Street, 79601 Lviv, Ukraine*

[2] *Poznań University of Technology, 13a Nieszawska Str., 61-022 Poznan, Poland*

**Abstract**

An original modification of the standard Pulse Laser Deposition (PLD) method for preparing both undoped and indium doped zinc oxide (ZnO:In) thin films at low substrate temperature is proposed. This preparation method does not demand any further post-deposition annealing treatment of the grown films. The developed method allows to grow thin films at low substrate temperature that prevents them from the considerable loss of their intrinsic electrical and optical properties. The influence of deposition parameters on the electrical and optical parameters of the undoped and the indium doped ZnO thin films is also analysed.

*Corresponding author: V.Savchuk
Ph.: + 380 (322) 65-19-84;
Fax: +380 (322) 63-70-88;
E-mail: vikosav@iapmm.lviv.ua

## 1. Introduction

A review of research works on luminescent materials shows that oxide materials with a wide energy band gap demonstrate a high quantum efficiency of luminescence [1-3]. The luminescent oxide materials can be an alternative for the traditional luminescent materials such as doped zinc sulphide (ZnS). In this context, oxide materials can be used as host materials for doping by rare-earth elements. The doped oxide thin films can be applied in electro- and cathodoluminescent displays as emitting medium with highly-effective emission [4, 5].

Pure zinc oxide (ZnO) is a wide band gap (3.2 eV at 300 K) semiconductor with a good conductivity. Moreover, this material is characterised by high transmittance in the visible spectrum region, high electric conductivity and good chemical stability in comparison with traditionally used commercial luminescent materials [4, 6-9]. These properties give opportunity for a wide range of technical applications of ZnO, especially in the form of thin films, in different opto- and microelectronic devices, for example, as transparent conducting electrodes in solar cells and various sensors. ZnO also has a great potential as light emitting material in the visible region, when doped with rare-earth atoms. Therefore, preparation of doped ZnO thin films opens new possibilities for using it as emitting medium in light emitting diodes, near ultraviolet and blue lasers.

Most publications deal with pure ZnO films or doped ZnO by rare-earth atoms [4-6, 9, 10-14]. Usually, the doped ZnO films with optimum properties (perfect crystalline structure, good conducting properties, high transparency, high intensity of luminescence) are obtained when they are grown on heated substrates and annealed after deposition at high temperature in oxygen atmosphere. However, for an extensive use in the commercial applications pure and doped ZnO films must be prepared at much lower substrate temperatures. Therefore, it is necessary to develop a low-temperature deposition technology for the growth of ZnO films. In comparison with the magnetron sputtering [1, 6, 9, 11,12, 14] and MBE methods [15-17], which are modern methods for the growth of thin films, Pulsed Laser Deposition (PLD) method [18-20] has many advantages and technological possibilities. These advantages are: effectiveness and simplicity of the deposition equipment, high deposition rates, wide spectrum of deposition parameters for the control and the optimisation of film properties, accurate control of stoichiometry and film thickness, use of background atmosphere of various gases, and high energy and high reactivity of the ablated material. The sum of all these special features enables the growth of oxide thin films at low temperature substrates with perfect crystallinity and good stoichiometry.

The present research is a continuation of our early studies of oxide luminescent films prepared by the PLD method [18-20], in which the influence of oxygen pressure, substrate temperature and concentration of indium atoms on electrical parameters of the grown films are investigated.

## 2. Experimental

The laboratory technological equipment use in the present work has been presented and described in our early article [5, 19, 20]. The beam with wavelength of 1064 nm, pulse duration of 15 ns and frequencies from 0.1 J/cm$^2$ to 1 J/cm$^2$ from a Q-switched Nd:YAG laser was used for ablating target materials. The ablated material was condensed in the reactive oxygen atmosphere on sapphire and quartz substrates at various temperatures and oxygen pressures. The PLD process occurred in a specially designed, quasi-closed reaction chamber (QCRC), which was placed inside the main vacuum chamber. The application of the QCRC allows use of oxygen as reactive atmosphere during PLD process. Simultaneously, it has allowed to prevent the contamination of growing films by the remainder gases of the main vacuum chamber. High purity oxygen (99.999%) was used in our experiments. During the PLD processes the targets rotated with a frequency of 10 Hz. The parameters of the growth process are listed in Table I.

The structural, electrical and optical properties of the grown films were investigated as a function of the deposition temperature and the oxygen pressure. The quality of the individual ZnO films was evaluated primarily by means of X-ray diffraction analysis. The resistivity of the grown films was measured by four-point probe. Electroluminescent parameters of the prepared thin film structures were measured at 50 microseconds bipolar square pulses with voltages of 10-50 V having different frequencies.

Table I. Parameters used for growth

| Growth parameter | Value or specification |
|---|---|
| Target | ZnO:In (ZnO) |
| Spot size on the target | ~2-3 mm |
| Energy density | Up to 1 J/cm$^2$ |
| Wavelength | 1064 nm |
| Pulse duration | 15 ns |
| Repetition rate | 12-56 Hz |
| Deposition rate | From 610 Å/min up to 200 Å/min |
| Target-substrate distance | 20-30 mm |
| Target rotation | 10 Hz |
| Pressure of the reactive gas | Up to 10 Torr |
| Substrate temperature | From 150 $^{o}$C up to 550 $^{o}$C |
| Thickness of grown films | 0.5-3.0 μm |

### 3. Results and discussion

The film growth rates were determined from the samples thickness, up to 1 μm. We established that the growth rates of ZnO films have strong dependence on the value of oxygen pressure. It has been observed that an increase of the oxygen pressure, in the QCRA, from $10^{-3}$ Torr to $10^1$ Torr causes a decrease in the growth rates from 610 Å/min up to 200 Å/min. Fig. 1 shows these growth rates as a function of oxygen pressure in the QCRA during the deposition process.

The lower growth rates were obtained when the total pressure of oxygen was increased. In the case when the oxygen pressure is high, the mean free path (MFP) (*l*) of the evaporated particles is short. During the expansion of the evaporated particles from the target surface to the substrate surface a large number of mutual collisions occurs between the evaporated particles and the background oxygen. Under such conditions, the MFP of the evaporated particles decreases proportionally to the increase of the oxygen pressure.

According to the molecular-kinetic theory of gases the MFP can be determined by the following formula:

$$l = \frac{kT}{\sqrt{2\pi d^2} P} \quad (1)$$

where *k* is Boltzman's gas constant, *T* is the room temperature in Kelvin's degree,

$d = 2.98 \times 10^{-10}$ m is the diameter of oxygen molecules, and $P$ is the pressure of the ambient oxygen. The MFP is calculated by using this formula as the function of the oxygen pressure:

$$l = 10^{-4}/P \text{ [m]}. \qquad (2)$$

When the oxygen pressure is lower than $10^{-3}$ Torr, the MFP is much larger than the target-substrate distance ($d_{st}$ = 25-30 mm). As a result, direct lines can describe the trajectory of the evaporated particles. In this case, scattering of the evaporated particles in the background oxygen is almost absent and deposition rates reach the value of 610-560 Å/min.

When the oxygen pressure is as high as $10^{-2}$ Torr, a large number of mutual collisions between evaporated material and background oxygen takes place. When the oxygen pressure is $5 \times 10^{-2}$ Torr, the MFP is much smaller than the target-substrate distance. Hence, the deposition rate is significantly reduced at an oxygen pressure higher than $10^{-2}$ Torr. The deposition rates move fastly down to 20 nm/min at such conditions. Moreover, the large number of mutual collisions causes significant reduction of the energy of the evaporated material during flight time from the target to the substrate. At the same time, the large number of the collisions of evaporated material with reactive background oxygen plays a key role in producing the oxide precursors required for the growth of the thin films with the stoichiometric compound.

The ZnO films grown at room temperature were amorphous. The increase in substrate temperature causes the improvement in the degree of crystalline perfection in the grown films. These films showed polycrystalline structures at increased substrate temperature, up to 400 $^0$C. An increase in the substrate temperature up to 450 $^0$C does not affect the polycrystalline nature of the films. The ZnO films deposited at elevated temperatures up to 480 $^0$C were textured.

X-ray diffraction analysis of ZnO films grown at 480 $^0$C shows a narrow and strong peak, which is located at 2θ = 34.82 (Fig. 2). This peak corresponds to the (002) orientation of the films. The peak located at 2θ = 42.15 is from the sapphire substrate. The analysis indicates that the grains are strongly oriented in the basal plane direction and were grown along the C-axis. The sharpening of the peak with increasing substrate temperature indicates that the grain size increased, also. The full width at half maximum (FWHM) of the (002) peak is as narrow as $0.16^0$. Therefore, the average grain size is estimated to be about 55 nm. In this case, the substrate temperature is the defining factor for the formation of high-degree crystalline perfection in the grown films.

All the films (ZnO and ZnO:In) prepared by the developed method, demonstrated excellent optical parameters and very good adhesion to the substrate surface. The transmittance of the ZnO films increases significantly on the substrate temperature, from room temperature to above 480 $^{0}$C. It was also discovered that the transmittance of the ZnO films increases with increasing oxygen pressure during the deposition process. The films prepared at optimum deposition conditions, i.e. when the substrate temperature is 480 $^{o}$C and the oxygen pressure $P = (4\text{-}8) \times 10^{-2}$ Torr, showed transmittance up to 85%. The measured transmittance of ZnO films grown at various deposition conditions is shown in Fig. 3. For comparison, this value is approximately equal to the transmittance of indium tin oxide.

The obtained results testify that both the oxygen pressure and the substrate temperature are key factors, which determine the crystalline perfection and the transmittance of the ZnO films. Moreover, the influence of the substrate temperature on the growth rate of the films obtained at the optimum of oxygen pressure ($P=(4\text{-}8)\times 10^{-2}$ Torr) is less noticeable than the influence of oxygen pressure on the growth rate of the films obtained at the optimal substrate temperature ($T = 480$ $^{0}$C).

The electrical parameters of the ZnO films prepared at various deposition conditions were investigated. The results revealed that the conductivity of the ZnO films depends on the values of both the substrate temperature and the oxygen pressure. We determined that the resistivity of the ZnO films increases with an increase of the oxygen pressure in QCRC during growth. The ZnO films grown at the substrate temperature of 480 $^{o}$C and the oxygen pressure lower than $10^{-3}$ Torr showed a low resistivity of $10^{-1}$ Ωcm. The resistivity of the ZnO films grown at the same substrate temperature, but at the oxygen pressure above $5 \times 10^{-2}$ Torr, achieved values greater than $10^5$ Ωcm. Fig. 4 shows the resistivity of the ZnO films, which have been grown at 350 $^{0}$C and 480 $^{0}$C, as a function of the oxygen pressure.

The low resistivity can be associated with considerable deficiency of oxygen and/or very large concentration of zinc interstitials. The oxygen vacancies and interstitial zinc atoms, in depending on relationship between them, can play a leading role in the electrical conductivity of the films. It is well known [21] that interstitial zinc atoms are the main factor, which induces native n-type conductivity in ZnO. Actually, undoped ZnO films demonstrates the n-type conductivity. These phenomena can be explained by the fact that a large number of interstitial zinc atoms in the film volume behaves as donors. For the samples deposited at 480 $^0$C and the oxygen pressure about $(6-8)\times10^{-2}$ Torr, the measured resistivity was greater than $10^5$ $\Omega$cm. Following these deposition parameters, the interstitial Zn atoms can be additionally oxidized, and incorporated into the film crystal lattice. Thus, the use of higher oxygen pressures leads to the growth of more transparent and more stoichiometric, but highly resistive, ZnO films.

As seen in Fig. 4, doping of pure ZnO films by indium atoms decreases the film resistivity by more than four orders of magnitude. The increase in the conductivity can be explained in terms of indium atoms which, at given deposition parameters, can take interstitial positions in the crystal lattice. In this case, indium atoms act as donors. Therefore, it is possible to make the conclusion that the values of resistivity of the ZnO:In films depend on oxygen pressure, growth temperature and concentration of indium atoms.

For observation electroluminescence (EL), the thin film structures were made. The sapphire plates, thickness of about 0.2 mm, previously covered by metallic electrodes, were used as the insulating layers. In this case, the aluminium layers, with thickness of 3 $\mu$m that were deposited on the sapphire sheets were used as a bottom electrode. Indium doped ZnO films were deposited on such substrate. Finally, of In-doped ZnO films were covered by ZnO film with low resistivity. The low-resistivity ZnO films played the role of top electrodes. Typical EL spectra observed and measured at room temperatures.

The EL parameters of manufactured structures with ZnO:In thin-film emitting layers exhibited strong dependence on both the deposition conditions and the concentration of doping atoms. By evaluating the EL parameters of the prepared structures, we found the optimal condition for the deposition conditions of ZnO:In films. The maximal intensity of EL can be obtained when the ZnO:In films were grown at the oxygen pressure $P = (6-8) \times 10^{-2}$ Torr and the substrate temperatures higher than 450 $^0$C.

Optimization of the chemical composition of the ZnO:In compound was carried out. In all case, the maximum value of EL spectra is located in suitably equal range of wavelengths of light emission for all concentration of doped atoms, which were examined. The effect of strong influence of the concentration of doped atoms on the EL intensity was observed.

The peaks of the EL intensity were located at 500 nm (0,5 mol% In doped), 505 nm (1 mol% In doped), 515 (2 mol% In doped), 520 nm (3,5 mol% In doped), 515 nm (5 mol% In doped). As a result, the following regularities were established:

- the EL intensity increases very rapidly, when the concentration of the doped atoms increases from 1 mol.% up to 3,5 mol.%;
- the EL intensity decreases fluently, when the concentration of the doped atoms increases from 3,5 mol.% to more 5 mol.%.

Figure 5 shows the EL spectra of prepared structures obtained at optimal deposition conditions as a function of the concentration of doping atoms.

The results of conducted investigations revealed that the maximum values in the EL spectra are located at nearly the same values of wavelengths for all the concentration of doping atoms. The maximal intensity of EL spectra for ZnO:In films changes depending on the contents Indium of atoms. But, on all cases, they were located the between 500 nm and 520 nm wavelengths, which closed the green band of visible spectra.

## 4      Conclusions

We have developed a new pulsed laser deposition method that uses oxygen as a reactive background gas during the deposition of oxide films. The developed method enables to prepare oxide films without any annealing treatment. ZnO films having a high transmittance, a good conductivity and a high chemical stability have been prepared with this method. These films can be used in applications in microelectronics and related high-technology applications. Moreover, the following features of the In-doped ZnO films were found:

(1) The increase in the substrate temperature above 460 $^{o}C$ improves the crystalline structure of the films and increases their transmittance. The films prepared at

480 $^{\circ}$C and oxygen pressures higher than $10^{-2}$ Torr show the transmittance up to 85%.

(2) The substrate temperature defines the crystalline structure of the films. The films grown at substrate temperature higher than 480 $^{\circ}$C were textured.

(3) The influence of oxygen pressure on the resistivity and transmittance of the grown films is an independent, but a very influential, factor. The indium doping of the ZnO films decreases film resistivity more than four orders of magnitude.

(4) The ZnO:In thin films show green luminescence, the intensity of which depends on the content of the doping atoms. The maximum electroluminescence emission occurs at 3.5 mol% concentration of In atoms in the film volume.

**Acknowledgements**   One of the authors (V.Savchuk) thanks the NATO Advanced Fellowships Programme and the National Administration at the Information Processing Centre (Warsaw, Poland) for partial financial support of the research.

**FIGURES**

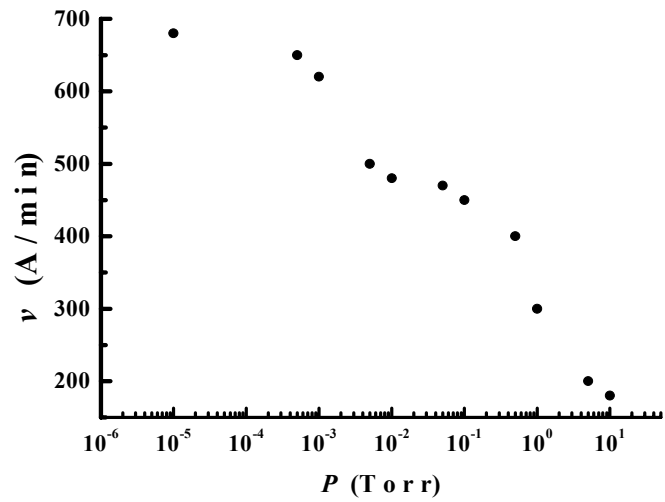

Fig. 1  Deposition rate of the ZnO thin films as function of oxygen pressure.

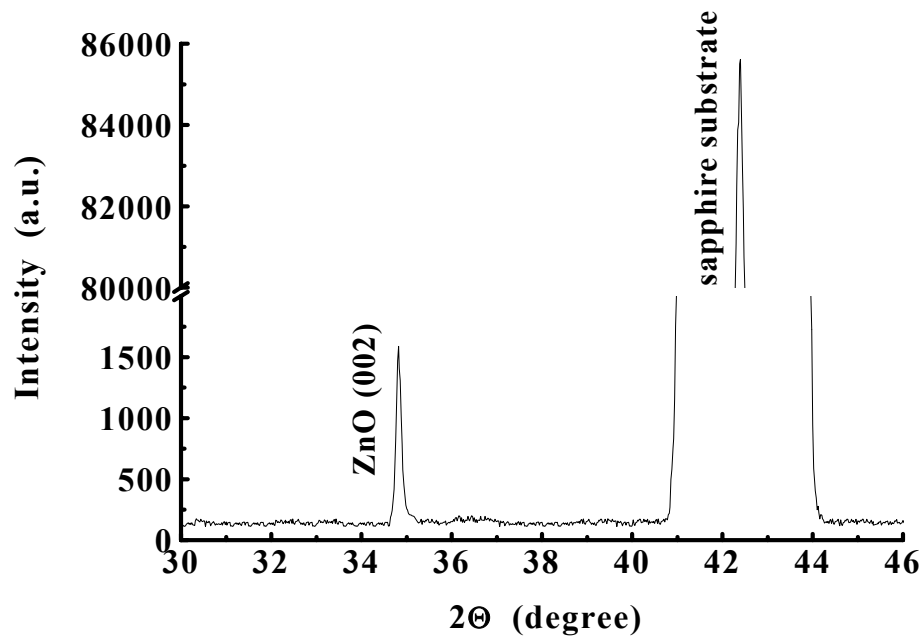

Fig. 2  X-ray spectra of ZnO film grown on Sapphire substrate. The film was deposited at optimal conditions (substrate temperature is 480 °C, oxygen pressure is $5 \times 10^{-2}$ Torr).

Fig. 3  Measured transmittance of ZnO films deposited on sapphire substrate as a function of oxygen pressure and substrate temperature.

Fig. 4  Influence of oxygen pressure on the resistivity of ZnO films prepared by PLD at different substrate temperatures.

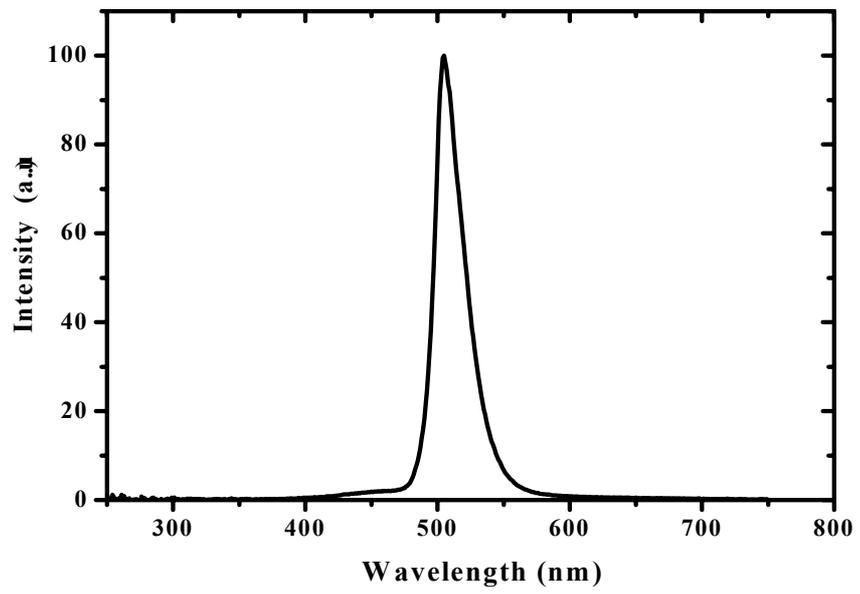

Fig. 5  Electroluminescence of a ZnO:In films deposited on sapphire substrate at 480 °C and $6 \times 10^{-2}$ Torr.